\documentclass[%
preprint,
aps,prl,floatfix,superscriptaddress,nopacs,footinbib,
amsmath,amssymb,
]{revtex4-1}

\usepackage{graphicx}
\usepackage{dcolumn}
\usepackage{bm}
\usepackage[normalem]{ulem}
\usepackage{hyperref}

\usepackage{color}
\newcommand{\bk}{\textbf{k}} \newcommand{\bq}{\textbf{q}}
 
\newcommand{\br}{\textbf{r}}

\newcommand{\mrm}{\mathrm}

\begin{document}


\title{Determination of the Superconducting Order Parameter from Defect Bound State Quasiparticle Interference}

\author{Shun Chi}
\affiliation{Department of Physics and Astronomy, University of British Columbia, Vancouver BC, Canada V6T 1Z1}
\affiliation{Stewart Blusson Quantum Matter Institute, University of British Columbia, Vancouver BC, Canada V6T 1Z4}
\author{W. N. Hardy}
\affiliation{Department of Physics and Astronomy, University of British Columbia, Vancouver BC, Canada V6T 1Z1}
\affiliation{Stewart Blusson Quantum Matter Institute, University of British Columbia, Vancouver BC, Canada V6T 1Z4}
\author{Ruixing Liang}
\affiliation{Department of Physics and Astronomy, University of British Columbia, Vancouver BC, Canada V6T 1Z1}
\affiliation{Stewart Blusson Quantum Matter Institute, University of British Columbia, Vancouver BC, Canada V6T 1Z4}
\author{P. Dosanjh}
\affiliation{Department of Physics and Astronomy, University of British Columbia, Vancouver BC, Canada V6T 1Z1}
\affiliation{Stewart Blusson Quantum Matter Institute, University of British Columbia, Vancouver BC, Canada V6T 1Z4}
\author{Peter Wahl}
\affiliation{SUPA, School of Physics and Astronomy, University of St. Andrews, North Haugh, St. Andrews, Fife, KY16 9SS, United Kingdom}
\affiliation{Max-Planck-Institut f\"ur Festk\"orperforschung, Heisenbergstr. 1, D-70569 Stuttgart, Germany}
\author{S. A. Burke}
\affiliation{Department of Physics and Astronomy, University of British Columbia, Vancouver BC, Canada V6T 1Z1}
\affiliation{Stewart Blusson Quantum Matter Institute, University of British Columbia, Vancouver BC, Canada V6T 1Z4}
\affiliation{Department of Chemistry, University of British Columbia, Vancouver BC, Canada V6T 1Z1}
\author{D. A. Bonn}
\email{bonn@phas.ubc.ca}
\affiliation{Department of Physics and Astronomy, University of British Columbia, Vancouver BC, Canada V6T 1Z1}
\affiliation{Stewart Blusson Quantum Matter Institute, University of British Columbia, Vancouver BC, Canada V6T 1Z4}

\date{\today}

\begin{abstract}
The superconducting order parameter is directly related to the pairing interaction, with the amplitude determined by the interaction strength, while the phase reflects the spatial structure of the interaction. However, given the large variety of materials and their rich physical properties within the iron-based high-Tc superconductors, the structure of the order parameter remains controversial in many cases. Here, we introduce Defect Bound State Quasi Particle Interference (DBS-QPI) as a new method to determine the superconducting order parameter. Using a low temperature scanning tunneling microscope, we image in-gap bound states in the stoichiometric iron-based superconductor LiFeAs and show that the bound states induced by defect scattering are formed from Bogoliubov quasiparticles that have significant spatial extent. Quasiparticle interference from these bound states has unique signatures from which one can determine the phase of the order parameter as well as the nature of the defect, i.e. whether it is better described as a magnetic vs a nonmagnetic scatterer. DBS-QPI provides an easy but general method to characterize the pairing symmetry of superconducting condensates.   
\end{abstract}

\maketitle



In superconductors, a pairing interaction binds electrons into Cooper pairs, condensing them into a coherent ground state with an order parameter $\Delta_{\bk} = |\Delta_{\bk}|e^{i\phi_{\bk}}$. Here, $|\Delta_{\bk}|$ is the magnitude of the superconducting gap and $\phi_{\bk}$ is the phase of the order parameter~\cite{BCStheory,hirschfeld2011gap,ScalapinoRMP}. Uncovering the \textit{complex} order parameter is key to a determination of the pairing mechanism. In particular, the phase factor $e^{i\phi_{\bk}}$ gives insight into how two electrons overcome Coulomb repulsion and bind together. For conventional superconductors, attractive interactions mediated by phonons result in an $s$-wave state in which $\Delta_{\bk}$ has the same sign everywhere in momentum space~\cite{BCStheory}. The $e^{i\phi_{\bk}}$ factor changes sign only at higher energies where Coulomb repulsion dominates, and the chance two electrons come close to each other is reduced~\cite{BCStheory}. In high-temperature cuprate superconductors, the electron-phonon interaction is believed to be too weak to be responsible for pairing~\cite{ScalapinoRMP}, and strong on-site Coulomb repulsion favours a $d$-wave order parameter with a sign-change (or phase shift by $\pi$) in momentum space~\cite{hirschfeld2011gap,ScalapinoRMP}. For the case of iron-based superconductors, which have multiple bands crossing the Fermi energy, the precise order parameters are still controversial and could differ between different compounds, but Coulomb repulsion may again favour a sign-changing order parameter~\cite{Hirschfeld2016Review}.  


Many methods have been explored for determining superconducting order parameters, however most are sensitive only to $|\Delta_{\bk}|$, which controls the gap in the density of states and cannot directly detect a sign-change. For order parameters with nodes, such as the $d$-wave state in the single-band cuprates, a sign change can be inferred from the angular dependence of the order parameter $|\Delta_{\bk}|$. These methods were remarkably successful in studies of cuprate and heavy fermion superconductors~\cite{Hardy1993Dwave,Allan2013CeCoIn5,Zhou2013CeCoIn5}. However, a definitive identification of the phase still required specialized experiments that were sensitive to the phase factor $e^{i\phi_{\bk}}$. In cuprate superconductors, whose sign change involves a broken rotational symmetry, this was achieved through measurements detecting the sign change associated with rotations by $90^\circ$, using tunnel junctions, or through the detection of half flux quanta~\cite{Kirtley1995symmetry,van_harlingen_phase-sensitive_1995}. In contrast to cuprates, most iron-based superconductors possess a nodeless order parameter, suggesting an $s$-wave pairing state. A vast amount of research has been undertaken to determine whether or not there is a sign change in the order parameter between different sheets of the Fermi surface, designated either as an $s_{\pm}$ or an $s_{++}$ order parameter. The lack of both broken rotational symmetry and absence of nodes mean that the techniques which have delivered definitive evidence for the pairing symmetry in the cuprate superconductors are inconclusive for the iron-based superconductors~\cite{hirschfeld2011gap}. 

A more broadly applicable method to probe the phase exploits quantum interference between the quasiparticle wavefunctions. Quasiparticle interference (QPI), measured via scanning tunneling microscopy (STM), detects the oscillating pattern resulting from interference of quasiparticles scattered by defects, and hence is sensitive to the phase difference between the initial and final states. This technique has been successfully applied to a number of cuprate superconductors\cite{hoffman_imaging_2002,kohsaka_how_2008}. To extract information about the phase of the order parameter, QPI due to scattering by vortices has been analyzed to detect signatures of the sign-changing order parameter~\cite{Hanaguri2009Dwave,Hanaguri2012,Fan2015}. However, application to the iron-based superconductors has encountered difficulties: the QPI intensities are located near Bragg peaks, making it hard to discern the true QPI signal~\cite{Hanaguri2010comment,Hanaguri2010commentReply}. Also vortices are spatially more extended in iron-based superconductors due to longer superconducting coherence lengths, which complicates comparison with theoretical calculations that assume point-like scattering potentials.\cite{HAEM2015} 

Here we will show that this difficulty can be overcome by studying QPI between well-defined Bogoliubov quasiparticles which inherit the phase of the order parameter~\cite{Balatsky2006review}. In-gap bound states, the excitations of Cooper pairs due to defect scattering, are excellent sources of Bogoliubov quasiparticles. Within the gap, the density of states of the clean superconductor is zero, but near defects there is a contribution from the Bogoliubov quasiparticles which make up the bound states. These bound states are confined to the vicinity of defects in real space, but have sufficient spatial extent that they can be relatively well localized in momentum space~\cite{ARPES_InGapState}. The screening of the defect potentials by these in-gap Bogoliubov quasiparticles leads to \textit{defect bound state QPI (DBS-QPI)}. 
Defects play a vital role in this distinct form of QPI, both as the source of the Bogoliubov quasiparticles and as the scattering centre that leads to interference effects.  Characterization of this DBS-QPI provides a direct phase-sensitive measurement of the superconducting order parameter.

We study DBS-QPI in LiFeAs, one of the stoichiometric iron-based superconductors, whose surface after cleaving is ideal for STM study~\cite{ChiPRL2012,Allan2012,Hanaguri2012}. By comparing experimental data with theoretical simulations that use a realistic band structure, we provide solid evidence for a sign change in the superconducting order parameter between the hole and electron bands. In turn, this significantly constrains the form of the pairing interaction.



\begin{figure*}[t]
\centering
\includegraphics[width=0.87\linewidth]{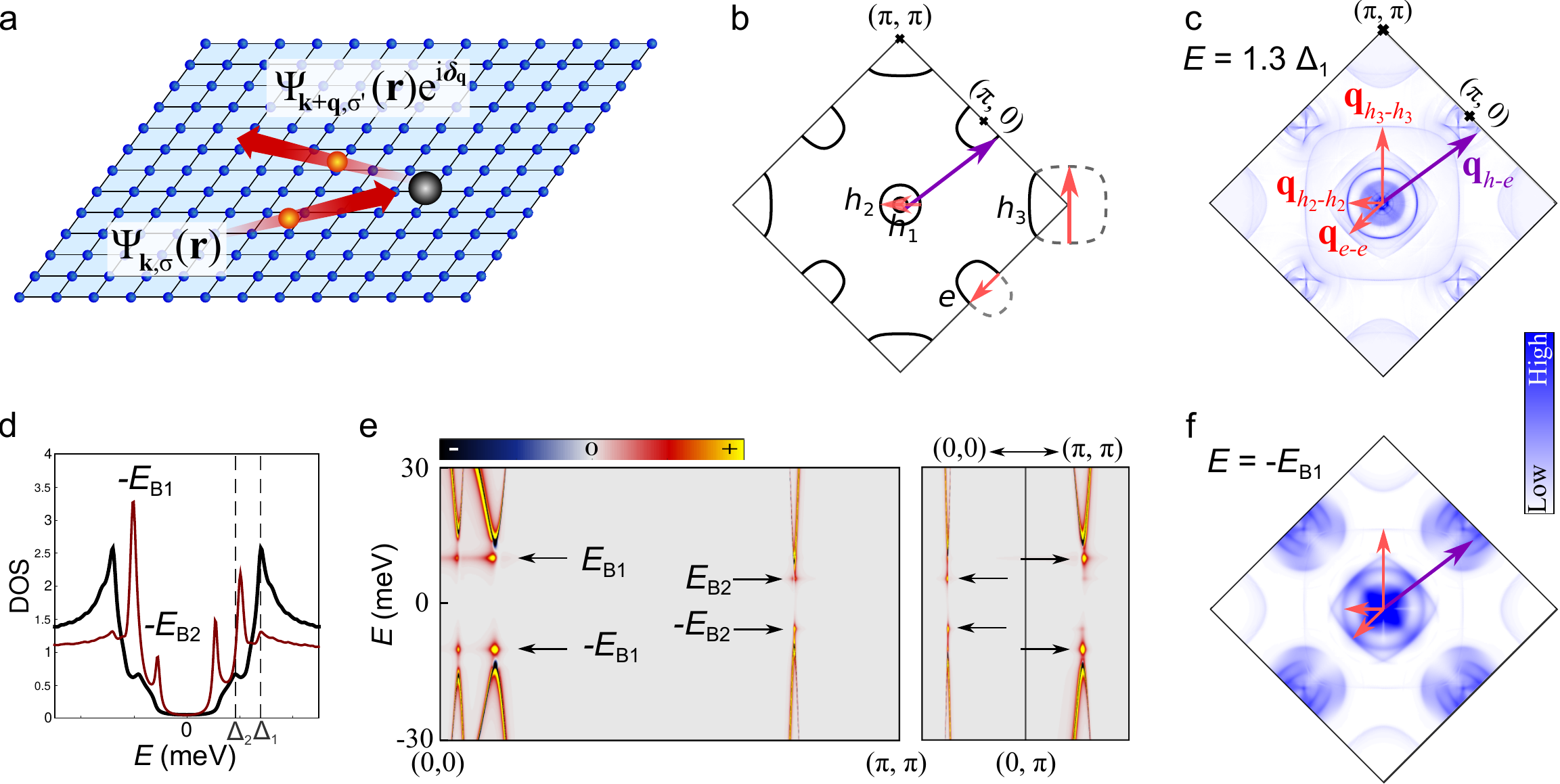}
\caption{\label{fig1} 
\textbf{Theory:} \textbf{a} Schematic of the quasiparticle scattering process.
\textbf{b} Fermi Surface of LiFeAs derived from a five-orbital model. The red and purple arrows indicate possible momentum transfers of the scattering process depicted in \textbf{a}). The three red arrows are intra-band scattering processes and the purple arrow is an inter-band scattering vector. \textbf{c} The simulated QPI for an energy above the superconducting gaps ($E = 1.3\Delta_{1}$) with the QPI features indicated by $\bq_{h_2-h_2}$, $\bq_{h_3-h_3}$, $\bq_{e-e}$, and $\bq_{h-e}$, respectively. $\bq_{h_1-h_1}$ corresponds to the ring-like feature in the center and is too small to be labeled. \textbf{d} LDOS for an $s_{\pm}$ order parameter: the unperturbed LDOS (black) shows two superconducting gaps, and the DOS on a nonmagnetic defect (red) gives two sets of in-gap bound states. \textbf{e} The defect-induced change of the DOS $\delta \rho (\textbf{k}, E)$ is shown in $k$-space, revealing the additional states due to defect bound states at $\pm E_\mathrm{B1,2}$. \textbf{f} Simulated magnitude of the DBS-QPI at $-E_\mathrm{B1}$, showing intra-band and inter-band QPI features.
}
\end{figure*}

Fig.~\ref{fig1}a schematically shows the scattering from a defect present in the lattice. A quasiparticle travels in the lattice in the state $\Psi_{\bk,\sigma}(\br)$, where $\bk$ and $\sigma$ are the momentum and the spin quantum numbers. When the quasiparticle encounters a defect, it scatters elastically with a certain probability to a final state $\Psi_{\bk+\bq,\sigma^{\prime}}(\br)e^{i\delta_{\bq}}$. The square modulus of the sum of the wavefunctions of all the possible scattering events produces spatial modulations in the local density of states (LDOS), which are referred to as QPI. The allowed wave vectors $\bq$ connect the available states at a given energy. In LiFeAs, three hole bands ($h_1, h_2, h_3$) and two electron bands ($e$) cross the Fermi energy. Four examples of $\bq$ at the Fermi energy are shown in Fig.~\ref{fig1}b, with the red and purple arrows indicating intra-band and inter-band scatterings, respectively. Fourier transformation (FT) of the real-space oscillations in the LDOS yields the QPI features in $\bq$-space, with maxima in the amplitude corresponding to the dominant $\bq$ vectors. Fig.~\ref{fig1}c shows the calculated QPI pattern for the superconducting state using a five-orbital model~\citep{GastiasoroPRB2013} and an $s_{\pm}$ order parameter with a nonmagnetic scattering potential~\cite{ChiPRB2017}. At $E = 1.3\Delta_1$, the QPI features are essentially identical to those in the normal state~\cite{ChiPRB2017}. The scattering vectors of Fig.~\ref{fig1}b which have a significant degree of nesting are easily identified as the more prominent QPI features in Fig.~\ref{fig1}c.

In QPI, the relative phase term $e^{i\delta_{\bq}}$ is primarily determined by two factors: the nature of the defect, nonmagnetic vs magnetic and weak vs unitary, which causes a phase shift during scattering; and the intrinsic phase difference between the states before and after scattering. 
In a superconductor, the spontaneous gauge symmetry breaking sets the phase of Cooper pairs in momentum space, which is the phase of the superconducting order parameter. 

Scatterers play a second role in a superconductor, providing a distinct form of QPI. The interplay of the superconducting order parameter and the nature of defects yields bound states inside the superconducting gaps~\cite{Balatsky2006review}. In the case of a conventional order parameter without a sign change, only a magnetic defect can give rise to in-gap bound states. With a sign-changing order parameter, both magnetic and nonmagnetic defects can induce in-gap bound states. These bound states are excitations of the superconducting ground state, namely Bogoliubov quasiparticles that are produced by defect scattering. Bogoliubov quasiparticles inherit the phases of Cooper pairs at different momentum states~\cite{Tinkham}. By studying the relative phase term $e^{i\delta_{\bq}}$ in QPI, one is able to decode both the superconducting order parameter and the nature of the defects.

Fig.~\ref{fig1}d shows the calculated LDOS of LiFeAs at a defect-free site (in black) and a nonmagnetic defect site (in red), assuming an $s_{\pm}$ order parameter. There are two sets of impurity bound states inside the gaps. To better resolve the origin of the bound states, the defect-induced change of the density of states (DOS) $\delta\rho(\bk,\omega)$ is shown in Fig.~\ref{fig1}e. The outer set at energies $\pm E_{\mathrm{B1}}$ are the in-gap bound states for the large gaps in bands $h_1, h_2$ and $e$, and the inner set at energies $\pm E_{\mathrm{B2}}$ are the in-gap bound states for the small gaps in band $h_3$ and $e$. In the cases of a magnetic defect with $s_{++}$ and $s_{\pm}$ order parameters, they yield very similar results (see section III.D of Ref.~\onlinecite{ChiPRB2017}). These states consist of Bogoliubov quasiparticles that are relatively localized in energy but follow the dispersion of the bandstructure near $E_{\mathrm{F}}$.  This means the wavefunctions of the superconducting bound states are delocalized in $\br$-space. Thus, defect-bound state QPI (DBS-QPI) can be generated from these bound states with scattering vectors similar to \textit{conventional} QPI from states above the gaps. Fig.~\ref{fig1}f shows the calculated magnitude of the DBS-QPI in $\bq$-space at $-E_{\mrm{B1}}$. While all DBS-QPI features are generally consistent with the QPI seen outside the superconducting gaps in Fig.~\ref{fig1}c, there are some important differences. First, the DBS-QPI features are broadened because the momentum distribution of the bound state is wider due to their confinement to the general vicinity of the defect in real space. Second, and more interestingly, the DBS-QPI features due to inter-band scattering processes $\bq_{h-e}$, indicated by the purple arrow, become significantly enhanced. This is an interference effect involving the interplay between the phase of the order parameter and the nature of the defect, and it is \textit{this} effect that can be used to identify the sign-changing nature of the order parameter. 

\begin{figure}[ht]
\centering
\includegraphics[width=0.44\linewidth]{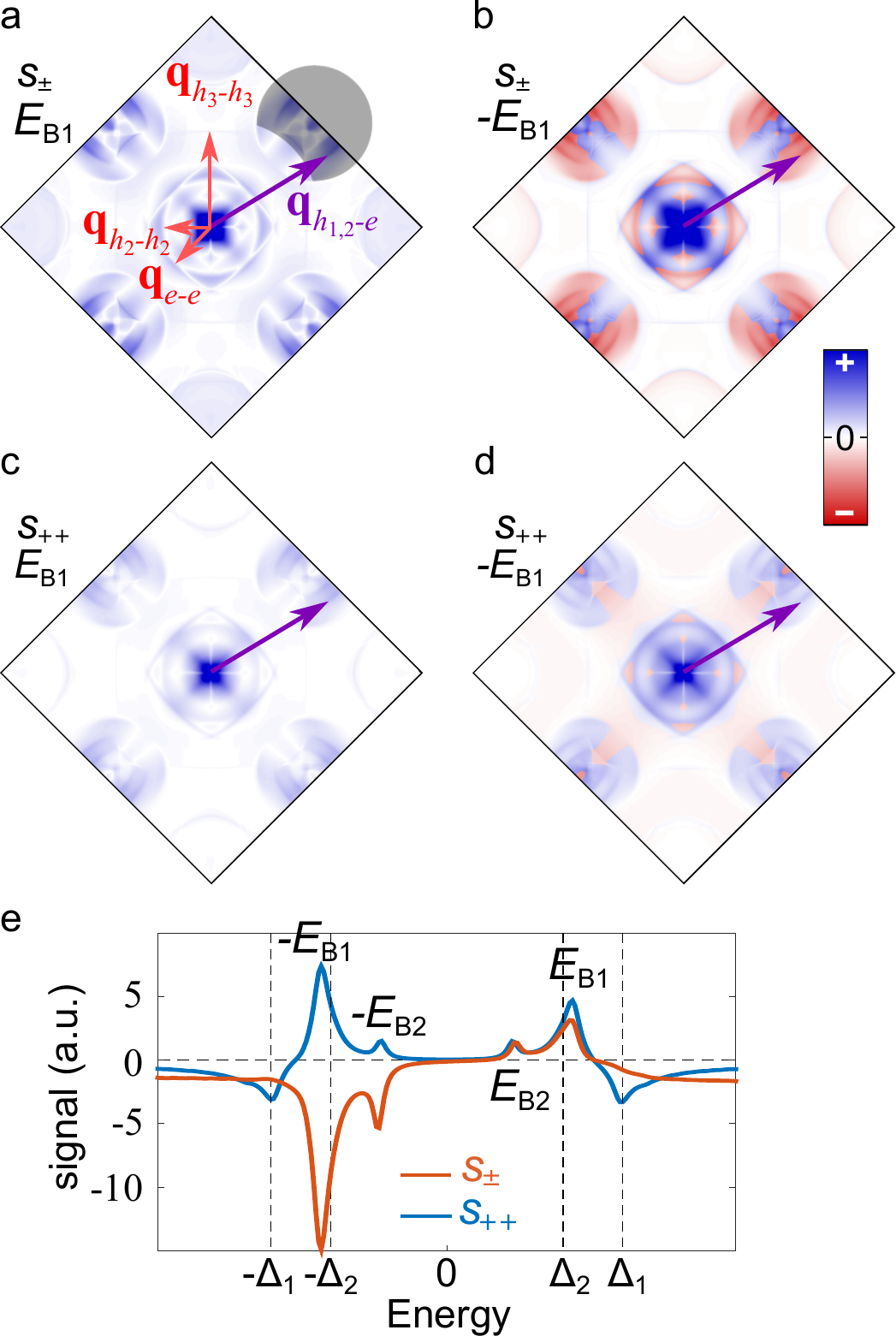}
\caption{\label{fig2} 
\textbf{Theoretical results of phase-referenced DBS-QPI.} \textbf{a, b, c,} and \textbf{d} The simulated phase-referenced DBS-QPI at $\pm E_{\mrm{B1}}$ for  $s_{\pm}$ with a nonmagnetic defect and $s_{++}$ with a magnetic defect, respectively. \textbf{e} The integrated $h$-$e$ DBS-QPI signal. The integration region is shown in \textbf{a} as a shaded area. 
}
\end{figure}
DBS-QPI distinguishes $s_{\pm}$ and $s_{++}$ through the phase of the Fourier transform. Here we define a \textit{phase-referenced} Fourier transformation (PRFT) to clarify this difference, which is at the basis of DBS-QPI. The experimentally measured quantity is the tunneling conductance map $g(\mathrm{\textbf{r}}, E)$, which is proportional to the LDOS $\rho(\mathrm{\textbf{r}}, E)$. 
The FT of $g(\mathrm{\textbf{r}}, E)$ is $|\tilde g(\mathrm{\textbf{q}}, E)|\times e^{i\theta_{\bq, E}}$, where $|\tilde g(\mathrm{\textbf{q}}, E)|$ is the intensity and $\theta_{\bq, E}$ is the phase of the Fourier component at wave vector $\bq$ and energy $E$. Conventionally, the phase is ignored, and only the intensity of the FT is analyzed. However, the phase is closely related to the scattering process and the interference of the quasiparticle wavefunctions. Analysis of the phase is complicated by the fact that it contains an arbitrary global phase factor related to the defect positions and details of defect apparent size and symmetry~\cite{ChiPRB2017}. Therefore, in order to extract meaningful information about the phase relation, we use the PRFT, which reveals the relative phase between QPI at positive and negative energies $\pm E$.  We extract the PRFT via the following steps: we first Fourier transform $g(\mathrm{\textbf{r}}, E)$ at positive energy $E$, obtaining the phase factor $e^{i\theta_{\bq, E}}$ which we use as reference for the Fourier transform at negative energy $-E$. The PRFT of the tunneling conductance $\tilde g_c(\mathrm{\textbf{q}}, \pm E)$ for $E>0$ is given by 
\begin{eqnarray}
\tilde g_c(\mathrm{\textbf{q}}, E) &=& |\tilde g(\mathrm{\textbf{q}}, E)|\times e^{i\theta_{\bq, E}}\times e^{-i\theta_{\bq, E}} \nonumber \\
&=& |\tilde g(\mathrm{\textbf{q}}, E)| \\
\tilde g_c(\mathrm{\textbf{q}}, -E) &=& \mathrm{Re}\ (|\tilde g(\mathrm{\textbf{q}}, -E)|\times e^{i\theta_{\bq, -E}}\times e^{-i\theta_{\bq, E}}) \nonumber \\
&=& |\tilde g(\mathrm{\textbf{q}}, -E)| \times \mathrm{Re}(e^{i(\theta_{\bq, -E}-\theta_{\bq, E})})
\label{g_c}
\end{eqnarray}
where Re means the real part. The phase factor $\mathrm{Re}(e^{i(\theta_{\bq, -E}-\theta_{\bq, E})})$ of the PRFT is $+1$ for in-phase oscillations, and $-1$ for out-of-phase oscillations. 

Fig.~\ref{fig2} shows the simulated phase-referenced DBS-QPI at $\pm E_{\mrm{B1}}$. The major DBS-QPI features seen here correspond to scattering vectors connecting bands with the large gaps, $\bq_{h_{1,2}-h_{1,2}}$, $\bq_{e-e}$,and $\bq_{h_{1,2}-e}$, as expected for the bound states associated with the large gaps. $\bq_{h_{3}-h_{3}}$ is still present because the small gap of the $h_3$ band is not fully open at $E_{\mrm{B}1}$ (see Fig.~\ref{fig1}d), but its strength is noticeably weaker than the other DBS-QPI features. The key difference between $s_{\pm}$ and $s_{++}$ is a sign change in the superconducting order parameter between hole and electron bands. Accordingly our analysis focuses on the inter-band DBS-QPI features, $\bq_{h_{1,2}-e}$, as indicated by purple arrows.  For $s_{\pm}$ with a nonmagnetic defect, the majority of the $\bq_{h_{1,2}-e}$ signal has the opposite sign between $\pm E_{\mrm{B1}}$, as shown in Fig.~\ref{fig2}a and \ref{fig2}b (blue for positive and red for negative). For the case of $s_{\pm}$ with a magnetic defect, the results are similar. In contrast, for $s_{++}$ with a magnetic defect, the sign of the $\bq_{h_{1,2}-e}$ signal is mostly the same between $\pm E_{\mrm{B1}}$, as shown in Fig.~\ref{fig2}c and \ref{fig2}d. The inter-band DBS-QPI features in $\tilde g_c(\mathrm{\textbf{q}}, E)$ is integrated over an area indicated in Fig.~\ref{fig2}a. The DBS-QPI signals as a function of energy is peaked at the bound state energies, as shown in Fig.~\ref{fig2}e. In particular, the signals show the opposite sign between the positive and negative bound state energies in the simulation with the $s_{\pm}$ order parameter. On the other hand, the simulation with the $s_{++}$ order parameter preserves the sign of the signal at $\pm E_{\mrm{B1}}$.


\begin{figure}[h!]
\centering
\includegraphics[width=0.44\linewidth]{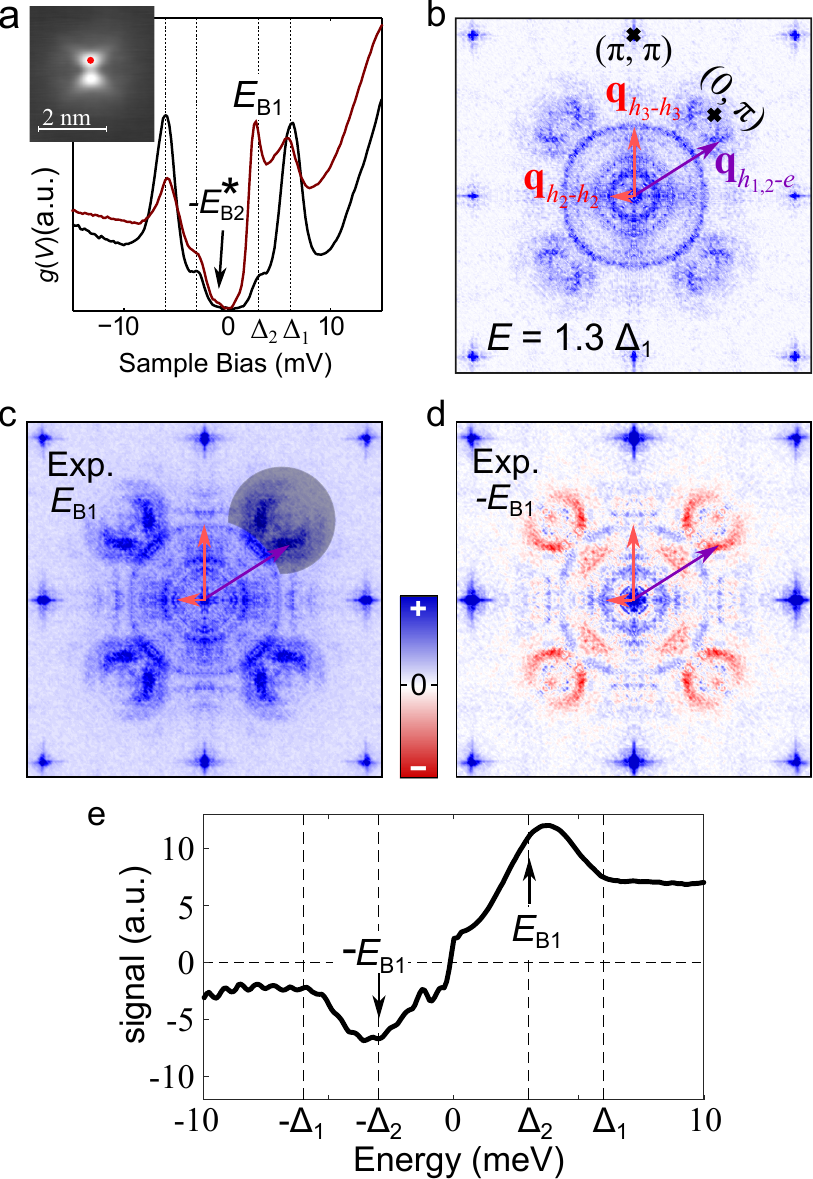}
\caption{\label{fig3} 
\textbf{Experimental results of phase-referenced DBS-QPI.} \textbf{a} The tunneling conductance $g(V)$ obtained on a clean area (black) and on an Fe-D$_2$ defect (red) ($T = 1.5$~K). The insert shows the topography of an Fe-D$_2$ defect ($V$ = 25 mV, $I$ = 50 pA). The red dot indicates the location for acquiring the spectrum. \textbf{b} QPI outside the gaps ($E = 1.3\Delta_1 = 7.8\ \mathrm{meV}$ ). The red arrows point to the intra-band DBS-QPI features, $\bq_{h_{1,2,3}-h_{1,2,3}}$, centered at $(0, 0)$. The purple arrow points to inter-band QPI features, $\bq_{h-e}$, centered at $(0, \pi)$.  \textbf{c} and \textbf{d} are the phase-referenced Fourier transform $\tilde g_c(\mathrm{\textbf{q}}, \pm E_{B1})$. \textbf{e} The integrated $h$-$e$ DBS-QPI signals from experimental data with the integration area indicated in \textbf{c}. Here sample bias (mV) is converted to energy (meV). 
}
\end{figure}

Next, we show the application of this technique to study the measured DBS-QPI for LiFeAs. Single crystals of LiFeAs ($T_c$ = 17.2 K) were grown 
using a self-flux technique \cite{ChiPRL2012}. For DBS-QPI measurements, a sample of LiFeAs was cleaved in-situ at a temperature below 20 K and inserted into a Createc scanning tunneling microscope (STM). DBS-QPI data were acquired at a base temperature of 4.2~K by taking $I$-$V$ spectra at each pixel in a grid of $400\times400$, and then numerically differentiating $I$-$V$ spectra to produce tunneling conductance maps. A home-built low temperature STM operating at temperatures down to 1.5~K was used to measure point spectra on native defects~\cite{white_stiff_2011}. The lower base temperature enables us to better resolve the bound states \textit{inside} the superconducting gaps, pinpointing the energies to focus on in the DBS-QPI analysis.
In as-grown LiFeAs, the most common native defect is the Fe-D$_{2}$ defect whose topography is shown in the insert of Fig.~\ref{fig3}a\cite{GrothePRB}. Measured at 1.5~K, the tunneling spectra taken at a defect-free area (black) show two superconducting gaps with $\Delta_1 = 6$~meV and $\Delta_{2} = 3$~meV, consistent with previous reports\cite{ChiPRL2012,Allan2012,Hanaguri2012}. The spectrum on an Fe-D$_{2}$ defect shows a prominent bound state inside the large gap and close to the edge of the small gap $E_{\mrm{B1}}\sim 3$ meV~\cite{GrothePRB,ChiPRB2016,chi_imaging_2017}. There is a shallow shoulder feature at $E^{*}_{\mrm{B2}} \sim 1.3$ meV inside the small gap, which is likely the inner set of bound states. However, unambiguous identification of this state for the FeD$_{2}$ defect requires higher energy resolution. Fig.~\ref{fig3}b shows QPI measured outside the gaps ($E = 1.3\Delta_1$). The QPI features in the experimental data agree very well with the simulation (see Fig.~\ref{fig1}c) except for the absence of $\bq_{e-e}$ which is consistent with previous reports~\cite{Allan2012,ChiPRB2014_QPI} and is probably due to a tunneling matrix effect. 

Fig.~\ref{fig3}c and \ref{fig3}d show the phase-referenced DBS-QPI at $\pm E_{\mrm{B1}} = \pm3$~meV. 
The Fe-D$_2$ bound state at $E_{B1}$ produces QPI that resmbles the features seen in Fig.~\ref{fig2}, confirming the states at $\pm E_{\mrm{B}1}$ are the in-gap bound states of the large gap. A sign inversion in the $\bq_{h_{1,2}-e}$ signal occurs between positive and negative bound state energies, as indicated by the purple arrows. This sign inversion is further confirmed by integrating the inter-band DBS-QPI (see Fig.~\ref{fig3}e) using the integration area indicated by the shadowed area in Fig.~\ref{fig3}c. Plotted as a function of energy in Fig.~\ref{fig3}e, the $\bq_{h_{1,2}-e}$ signal peaks at the bound state energies $\pm E_{\mrm{B}1}$ but with the opposite signs at the two energies. This is only consistent with the simulation using the $s_{\pm}$ order parameter (see Fig.~\ref{fig2}). In the experimental data, the absence of features at $\pm E_{\mrm{B}2}$ in the integrated signals is due to thermal broadening at the temperature of the measurement (4.2K) and the weakness of the signal compared to measurement noise.

\begin{figure*}[t]
\centering
\includegraphics[width=0.87\linewidth]{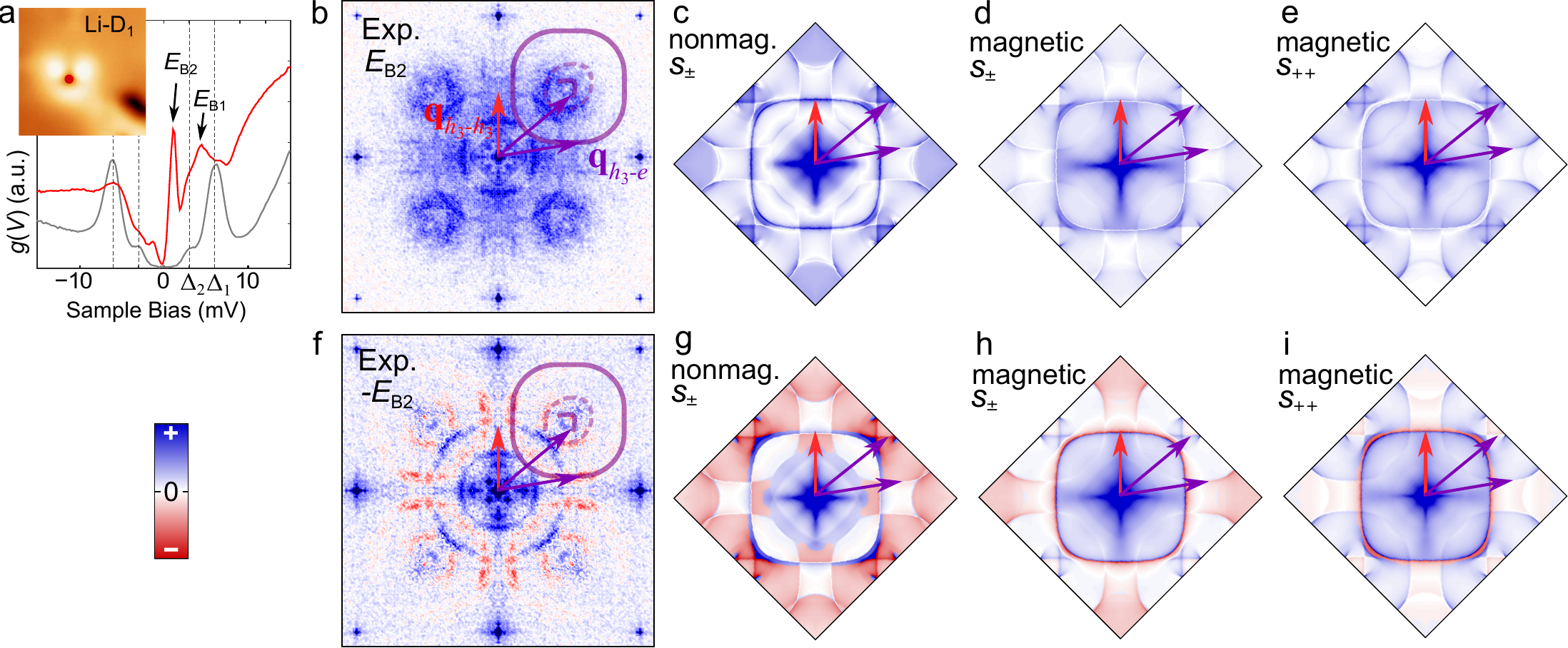}
\caption{\label{fig6} 
\textbf{Phase-referenced DBS-QPI associated with the small gaps.} \textbf{a} $g(V)$ measured on a Li-D$_1$ defect shows two sets of bound states. The strong bound state at $\pm E_{\mrm{B2}}$ is associated with the small gaps. \textbf{b} and \textbf{f} Measured phase-referenced DBS-QPI at $\pm E_{\mrm{B2}} = \pm 1.2~\mathrm{meV}$. Three QPI features centered at $(0, \pi)$ are highlighted by the arc and ovals, two of which are $\bq_{h_3-e}$ indicated by purple arrows. The position of $\bq_{h_3-h_3}$ is also specified by a red arrow. \textbf{c, d, e, g, h} and \textbf{i} Calculated phase-referenced DBS-QPI for the three cases allowing in-gap bound states.
}
\end{figure*}


In addition to Fe-D$_2$ defects, a few other native defects are present and give strong bound states inside the small gap, for example the Li-D$_1$ defect whose $g(V)$ is shown in Fig.~\ref{fig6}a. The phase-referenced DBS-QPI was measured at $E_{\mrm{B2}}\sim 1.2$~meV for the bound state associated with the small gaps, and is shown in Fig.~\ref{fig6}b and \ref{fig6}f. Three inter-band QPI signals dominate, as indicated by the purple shapes. The middle oval (dashed shape) is the $\bq_{h_{1,2}-e}$ QPI feature from $E_{B1}$ and present here because of thermal broadening. The other two shapes are QPI features from scattering between in-gap bound states for the small gaps in $h_3$ and $e$ bands, $\bq_{h_{3}-e}$. Consistent with the results above for $\bq_{h_{1,2}-e}$, $\bq_{h_{3}-e}$ has a sign inversion between $E_{\mrm{B2}}$ and $-E_{\mrm{B2}}$
whereas QPI features for intra-band $\bq_{h_3-h_3}$ scattering preserve the sign. The calculated phase-referenced DBS-QPI are shown in Fig.~\ref{fig6}c-d and \ref{fig6}g-i for the three possible scenarios that allow in-gap bound states: $s_{\pm}$ with a nonmagnetic defect, $s_{\pm}$ with a magnetic defect, and $s_{++}$ with a magnetic defect. With $s_{\pm}$ and a nonmagnetic defect, the $\bq_{h_3-e}$ signal changes sign and the $\bq_{h_3-h_3}$ signal retains the same sign between $E_{\mrm{B2}}$ and $-E_{\mrm{B2}}$. For the other two cases with a magnetic defect, the $\bq_{h_3-h_3}$ features dominates the signal and changes sign between $E_{\mrm{B2}}$ and $-E_{\mrm{B2}}$, as indicated by the red arrows. The one that is in best agreement with the experimental data is the calculation using $s_{\pm}$ with a nonmagnetic defect. Hence QPI at the in-gap bound state for the small gaps further verifies the consistency between experiment and theoretical results assuming an $s_{\pm}$ order parameter. In addition, it identifies the nonmagnetic nature of the native defects in LiFeAs since none of the simulations with a magnetic defect fit the phase-referenced DBS-QPI at $E_{\mrm{B2}}$. 


In LiFeAs, the sign change of the superconducting order parameter between hole and electron bands indicates that electrons pair together between next-nearest-neighbor (NNN) sites. The most plausible interaction that is able to generate an attractive channel between NNN sites is due to stripe antiferromagnetic spin fluctuations~\cite{Mazin2008Theory,hirschfeld2011gap,ScalapinoRMP}. 

LiFeAs is not the only superconducting compound which shows DBS-QPI. The anti-phase oscillations had been predicted from theory for $d$-wave superconductors~\cite{Balatsky2006review} and seen in experimental observations of the bound states in cuprate and heavy fermion superconductors~\cite{Hudson2001Ni_defect,Zhou2013CeCoIn5}. DBS-QPI provides a robust method for revealing both the order parameter and the nature of defects in superconductors with unconventional order parameters. Given that the typical apparent size of a defect is only a few lattice constants, a $\delta$-function potential is a good approximation for theoretical modeling, which makes it easier to compare experiment with theory than when using vortex cores. However, the analysis of experimental data containing contributions to phase from multiple defects of finite size and non-point symmetry is aided here by the use of phase-referenced Fourier transforms to help isolate phase changes due to the order parameter.
\section{Acknowledgements}
The authors are grateful for helpful conversations with George Sawatzky, Mona Berciu, Peter Hirschfeld, Andreas Kreisel, and Steven Johnston. Research at UBC was supported by the Natural Sciences and Engineering Research Council, the Canadian Institute for Advanced Research, and the Canadian Foundation for Innovation. SAB was further supported by the Canada Research Chairs program. PW acknowledges support from EPSRC grant no EP/I031014/1 and DFG SPP1458.

\bibliographystyle{apsrev4-1}
\label{Bibliography}
\bibliography{LiFeAs_BoundStateQPI}
\end{document}